\documentclass[runningheads]{llncs}
\usepackage[utf8]{inputenc}
\usepackage[hyphens]{url}
\usepackage{xcolor}
\usepackage{graphicx}
\def\etal{\emph{et al. }}
\usepackage{flushend}
\usepackage{geometry}
\usepackage{wrapfig}
\usepackage{url}

\begin{document}

\title{\textbf{\emph{Caveat Venditor}, Used USB Drive Owner}}
\author{James Conacher, Karen Renaud, Jacques Ophoff\\Abertay University, Dundee}
\institute{cyber4humans@gmail.com}
\date{}

\maketitle

\section*{Abstract}
USB drives are a great way of transferring and backing up files. The problem is that they are easily lost, and users do not understand how to secure or properly erase them. When used to store private and sensitive information, this constitutes a risk that users may be unaware of. Consider that people sell used USB drives online  --- presumably either their own or drives others have lost. This raises some interesting questions, such as whether sellers know how to ensure that private data is erased before they relinquish the drive to an unknown buyer, and whether sellers use these drives in an attempt to compromise an unwary buyer's device. Governments do indeed issue advice about the risks of used mobile media, but we do not yet know whether this advice is reaching, and being heeded by, the general public. 
To assess the situation, a sample of used USB drives were purchased from eBay sellers to determine, first hand, what  was  on the drives. This acts as an indicator of \emph{actual} security-related behaviours to  answer the questions posed above.
Using forensic analysis, it was found that a great deal of private and sensitive information remained on many of the drives, but there was no trace of malicious software.
More effective ways of enlightening the public are needed, so that private data is not unwittingly leaked via sold used media.

\section*{Keywords}
Used Mobile Media, Digital Forensics, File Recovery, Privacy Loss, Security Awareness



\section{Introduction}

It is  important for the man and woman in the street to know how to keep their own information secure \cite{howe2012psychology,wash2010folk,norton}. They need to be aware of: (1) the \emph{need} for good security practice, and (2) \emph{how} to implement these practices. So far, drives to improve cyber security have mostly been push-oriented. The Scottish and UK governments, for example, do their best to  educate the public about the importance of good digital hygiene  \cite{10steps} by disseminating advice through websites \cite{govtadvice}. 

Of particular interest to this study is that the selling of used USB drives is discouraged \cite{10steps}. There are two reasons for this admonition. \emph{Firstly}, it is easy for sellers to unwittingly leave personal and sensitive information on the drives \cite{nhs}, which could easily compromise their privacy. 
\emph{Secondly}, the recipient of the drive is also at risk. People are advised not to plug someone else's used drive into their computers \cite{bbc}.
There are good reasons for this. One of the most well-known cyber exploits, Stuxnet, was installed via a USB drive \cite{langner2011stuxnet}. A number of cases of people  deliberately leaving drives lying around  to entice usage \cite{Schneier} indicate that this is not an unlikely attack vector. On the other hand, sellers might also unknowingly sell a drive with a virus or malware on it \cite{Mimoso}. Regardless of intention, the buyer's device will be infected when they use their new purchase.

The problem is that no one has a good idea of whether the general public is aware of these risks. 
One could ask them, but self-report is problematic in this context, as in others when asking people about behaviours where they know what the correct response \emph{ought} to be \cite{junger1999self,cyders2012relationship}. Even if people report their knowledge accurately and we find that they do indeed know about the risks, we are not accurately measuring their actual selling behaviours, nor whether they effectively ameliorate their risks if they engage in such behaviours.  
A more reliable way to measure behaviours reflecting security stances is to observe actions.
In other words, we could take a snapshot of  measurable  behaviours with security implications, and these could arguably act as an indicator of risk awareness and understanding \cite{storer2010investigating,grispos2013using,glisson2011electronic}. 

This study proposes to take such a snapshot by engaging with the phenomenon of used USB drives being sold on eBay. This allows us to snapshot actual risky behaviours, as an indicator of likely digital hygiene behaviours of the populace at large. The questions we seek to answer are:


\textbf{ Q1 --- Seller Risk:}  \emph{Do people leave private and sensitive information on  USB drives they relinquish voluntarily? }   
 
\textbf{ Q2 --- Buyer Risk:}   \emph{Do used drives contain malicious software that could compromise unwary buyers' devices?} 

We first review the related work in Section \ref{related}. We then discuss the ethical considerations of the study  (Section \ref{forensics}). In Section \ref{investigate}, we detail our investigation methodology. Section \ref{findings} reports on the outcome of our investigation, Section \ref{discuss} discusses and reflects on the findings and Section \ref{conc} concludes. 


\section{Related Work}\label{related}

\begin{figure}[ht]
\centering
  \includegraphics[width=0.8\columnwidth]{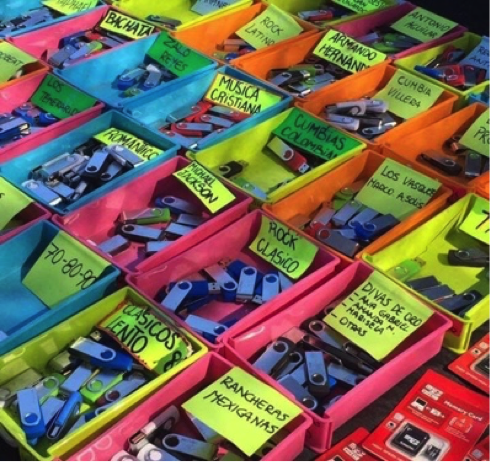}
  \caption{Music being sold in Mexico \url{https://www.reddit.com/r/Cyberpunk/comments/54oo49/music_being_sold_on_the_street_in_mexico_xpost/}}~\label{fig:mexico}
\end{figure}

 Figure \ref{fig:mexico} anecdotally reflects the widespread use of these  USB ``thumb'' drives.

Soon after USB drives started being widely used their risks became apparent. These risks apply to both  first and subsequent owners. 
In this section, we will discuss:
(1) people storing personal and sensitive information on USB drives, 
   (2)  unscrupulous people planting `Trojan' drives, specifically to target the unwary  or na\"{i}ve, and
 (3) people  finding such drives being likely to plug them in. 
Having established the risks to both buyer and seller of removable storage media, we then consider: 
 (4) the risks USB drive sellers run in selling their used drives.
 
We conclude by motivating this study to measure the incidence of this particular risky behaviour in the UK.

\paragraph{2.1 Personal and Sensitive Information on Drives \\}\label{sens}
The UK's National Health Service (NHS) London released statistics covering April 2008 to April 2009, during which 76 serious incidents  involving patients' and employees' personal information being lost or stolen had occurred.  They 
reported that 37,872 patients might have been affected  \cite{BBC1}. The NHS was fined £375k in 2012 for not ensuring that used drives were properly destroyed. The contractor they paid to dispose of their drives had  sold them online, without scrubbing them. 

In 2016, a study   by security software firm ESET reported that 22,000 USB sticks were found by dry cleaners in the UK every year
\cite{Schrott}. 
In 2017, 
a USB drive found in the street contained  details of the measures used to protect the Queen at Heathrow airport.  The drive was not encrypted and  some of the files were marked as ``Confidential'' or ``Restricted'' \cite{BBC1}.
In 2018, Rochester Grammar School lost
an unencrypted  drive containing the personal details of more than 1,000 pupils. 
There have also been a number of cases of USB sticks being sent by mail and being lost on en route \cite{HIPAA16}. 
%
%
Some large organisations, such as International Business Machines (IBM) and the NHS, now  discourage or ban the use of USB drives altogether \cite{Condon09,bbc2Lee}. 

\paragraph{2.2 Trojan Horse USB Drives \\}\label{trojan}
Plugging in a used USB drive is risky.  This is still true \cite{Nohl14,Ducky,Killer,Seals} despite the fact that computers no longer automatically execute the {\tt autoexec.bat} file when the drive is plugged in \cite{falliere2011w32}. 

USB baiting is the practice of dropping drives for people to find and plug in \cite{drop} to hack into a company's computer systems \cite{russians,Sterling}.
During the  Kim Jong-Un and Donald Trump summit,  USB-powered personal fans were handed out to the press.  Reporters were warned not to plug them in because  engineered USB devices could circumvent security measures and install malicious software \cite{BBC4}. 

\paragraph{2.3 People Plug Drives In\\} \label{plugin} People tend to plug in USB drives they find  \cite{Nichols16,stasiukonis2006social}. 
   A study carried out by the University of Illinois, Urbana-Champaign was a real-life study into  what members of the public would do with a `lost' USB drive found lying on the ground \cite{tischer2016users}.  300 drives were dropped around the campus, some with enticing labels such as ``Confidential'', or ``Final Exam Results'' or attached to a bunch of keys.   98\% of the  drives were picked up.  Of these, 45\% (135 devices) were  plugged into a PC, and at least one file  accessed. 
 
\paragraph{2.4 Risks of Used Drive Sales\\}\label{selling}
   People do indeed sell used USB drives via auction sites, such as eBay. A search on the 5th October 2018 displayed 98 auctions.
      Sellers could lose private and sensitive information they have  left on the drive. If the buyer abuses this information, the seller's privacy has been compromised. 
      
      Do people actually leave information on sold drives?
    Australian researchers \cite{robins2017investigation} bought used USB  drives from the Australian eBay site. The researchers wanted to see if any traces of data left on the devices remained, or whether they had been securely wiped.  They  analysed all the files to detect sensitive information left by the previous owner.   7.72\% of the drives contained no data, 1.11\% revealed no attempts to delete data, and 22.05\% had attempted, but failed, to delete the data on the drive.
   
   Other researchers \cite{storer2010investigating} bought second-hand mobile phone handsets  from eBay UK.   
They found less sensitive data than anticipated but  
they did find evidence  of images (including nudity and explicit material), pornography, drug use, bank account details, and 
personal health information.
This study concluded that further follow-up analysis of results was needed.

\paragraph{2.5 Summary \& Study Motivation\\}\label{summary}
The media reports on USB-related events  cannot reasonably be considered to be representative of the general public's security hygiene stance or behaviours --- most of the drives were lost, not sold. There might be many more losses that have been ``dark'', i.e. cases in which loss is not noticed or reported. The prevalence of harm might be more significant than media reports suggest. 

However, a study of devices or mobile media that are deliberately relinquished can deliver better insights into deliberate behaviours, including pointing out those that can be impacted by awareness and education drives. Studying this is, in many ways, a best-case scenario because  sellers know that they are giving access to unknown ``others''. One imagines that sellers have the opportunity and motivation to wipe a drive before selling it. Nevertheless there have been reports of devices with embedded hard drives being sold, with the seller unwittingly also giving the buyer access to information they ought not to have \cite{nhs,Varner18}.

The content of used (sold) drives might give a more accurate  indication of extant \emph{deliberate} security behaviours. The Australian study carried out by Robins \etal \cite{robins2017investigation} gives an indication of Australians'  USB drive related behaviours. This study aims to measure UK behaviours in this respect, because different  governments' awareness-raising and advice-issuing efforts are bound to have different levels of success.

\section{Ethical Considerations}\label{forensics}
Any forensics investigation has ethical considerations, and this study was no exception. The InfoSec code of ethics for forensics \cite{ethics1} instructs investigators to ``\emph{abide by the highest moral and ethical standards, and comply with all legal orders of the courts. To do so, he or she thoroughly examines all evidence within the scope of an investigation}.'' Moreover, and pertinent to this investigation, a forensics examiner will ``\emph{never reveal any confidential matters}''. Moreover, the following issues were considered.

\textbf{Ethical Concerns:} It should be noted that this was a research project and, as such,  not part of a criminal investigation. 
Even so, the
 forensics investigation had the potential to uncover very personal information about the sellers, but it might also contain unsavoury or illegal images.   We applied for ethical approval  to protect the privacy of the sellers, who were unwitting participants in our study. We also wanted to ensure that the investigator was not negatively affected by the contents of any files that might be found. 
 
\textbf{Ethics Approval:} Our ethics board approved the study with some important provisos. We were permitted only to reveal the file names, but not to open the files themselves. This was mandated  to ensure privacy protection of the seller, and to avoid traumatising the investigator. Hence our study reports only on the kinds of files found, and on inferences we were able to draw from the file names. 

\textbf{Platform Terms \& Conditions:} As the bulk of the buying and selling for this research project happened on eBay’s auction site, the User Agreement for the site was thoroughly checked to ensure that the proposal did not breach their Terms and Conditions.  We did not find any restrictions on  using the auction system to support research projects of this kind.
 
\textbf{UK Law:} Whilst the ultimate goal of this project is not to process any personal data, and safeguards are in place both to protect the researcher and members of the public, data that is inadvertently processed within this research project is done so under auspices of the Data Protection Act 2018, Schedule 1 Part 2 Clause 6(1)(b): ``is necessary for reasons of substantial public interest'' (United Kingdom., 2018b, sec. Schedule 1 Part 2 Point 6(1)(b)), and under The Data Protections Act 2018, Schedule 1 Part 2 Clause 6(2)(a) \cite{UKgovt}.

\textbf{European Law:} There is a requirement to hold all research data within the European Union (EU), in  line with the Global Data Protections Regulations, as enacted within the United Kingdom’s Data Protection Act 2018. We adhered to this requirement. 

\section{Investigation Methodology}\label{investigate}

We investigated general awareness and actions related to the use and protection of USB drives in two stages: first using a public survey, and second a forensic investigation. 

\paragraph{4.1 Survey \\}\label{survey}
This study  commenced with a public web-based survey.  The aim was to gain insights into general  USB-related risk awareness.  Participants were recruited using opportunistic snowball sampling. The survey was worded using appreciative inquiry \cite{cooperrider1987appreciative} to minimise social desirability responses and included attention check questions to weed out nonsense responses. 

The Microsoft Forms platform was selected to host the survey because it provides assurance that customers located within the EU had their responses stored within a data centre within the EU. A publicly accessible website displayed an introduction page providing context and a  link to the survey.
The website was configured to prevent search engine  indexing. 
It was hosted by GoDaddy within their European zone. 
In addition to this, the website was configured with a Security Certificate,  signed by a trusted Public Certification Authority, and McAfee Secure configured on the site  to maximise trust.

\paragraph{4.2 Ethical Forensics Methodology \\}\label{buying}
USB drives were purchased from a personal eBay account: one that was not associated with the host University.  The sellers became participants within the research study by virtue of selling a used USB drive on the site.  A total of 122 drives were purchased in order to ensure that at least 100 working drives remained to support analysis.  
Forensic investigations require maintenance of an audit trail. However, the participants' anonymity also had to be assured, so we constructed a scheme to ensure such anonymisation. Each drive was assigned a unique 3-digit number, randomly generated between 100 and 999, which which the drives were tagged. Mapping information (purchased drive:id number) was retained within an encrypted  spreadsheet stored  on an independent device. In a further step to anonymise the drives their order in the spreadsheet was randomised, to make it harder to link any drive to the previous owner.

\begin{figure}[ht]
\centering
  \includegraphics[width=0.5\columnwidth]{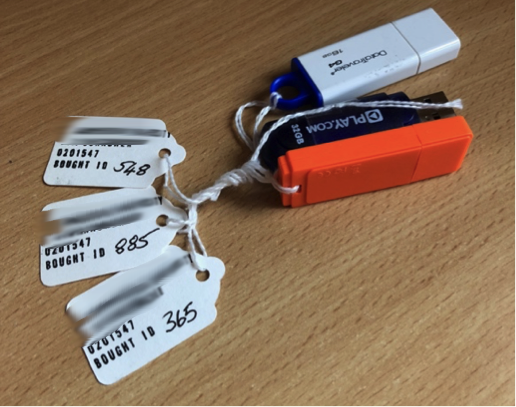}
  \caption{USB drives with anonymous labelling in Forensic Workstation setup.}\label{fig:tagging}
\end{figure}

Given the requirements of the ethical review board, and the InfoSec guidelines \cite{bassett2006computer} we formulated the following methodology, following as closely as possible the methodologies reported by \cite{jones2009,robins2017investigation,szewczyk20122012}.

\begin{figure}[h]
\centering
  \includegraphics[width=0.5\columnwidth]{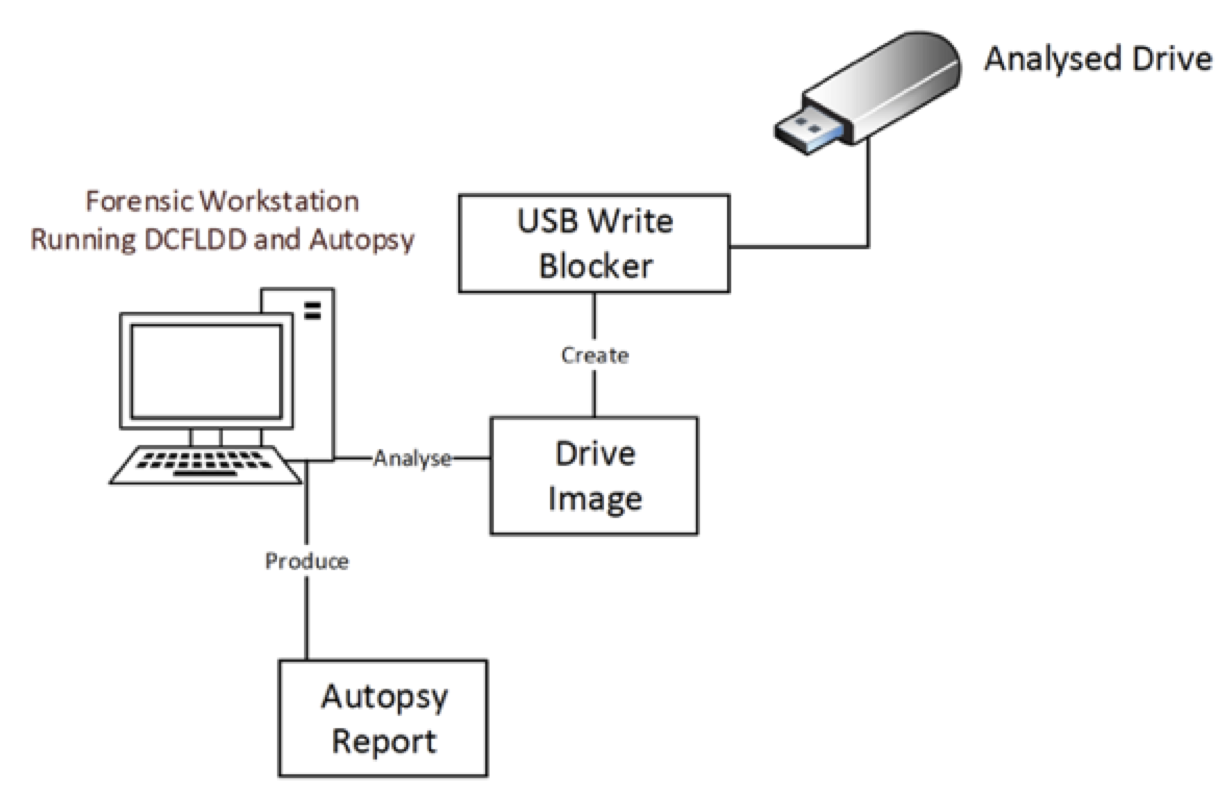}
  \caption{Forensic Workstation Setup}\label{fig:setup}
\end{figure}

\textbf{Investigation Hardware:}
Testing was carried out within a Univerity `Hacking Lab' (see Figure \ref{fig:setup}). The lab's computers are kept fully updated with current virus scanners installed and activated.  
        The computers are re-imaged every day 
         to ensure a clean operating system environment.  
         The computer that was used for the investigation was 
        disconnected from the  network ensuring  a completely standalone setup with no external tampering or leakage of drive contents possible.  This was done to ensure that any malware that was installed could not infect other computers.
        
      We used a USB write blocker to ensure that no data was changed on the source drive, and that the drive would be handled in a forensically-sound manner.   
     The Write Block device was validated using CRU WriteBlocking validation software.


\textbf{Pre-Analysis:}
\emph{(1) Visual Inspection:} Drives are  inspected visually to ensure that they did not appear to have been tampered with.  Any drive that is suspect is eliminated from the investigation.  (Tampering could include cases of drives that have been opened, or  traces of glue on the drive.)

\emph{(2) Protect:} Each  drive is inserted  into  a USB Write Block device.
     
\emph{(3) Plug In:} The inspected drive, within the write blocking device,  is  plugged into a dedicated computer within a protected University's Hacking Laboratory.
        If the USB drive fails to mount, or can not be read, it is added to the discard pile.

\emph{(4) Virus Scan:} Each drive is scanned for viruses or malware. 

\emph{(5) Create Image:} A forensically-sound image is created using freely available forensic imaging tools, mainly `DCFLDD'. This image would support forensic analysis.  Using DCFLDD, instead of DD, allowed us to calculate the MD5 checksum for the image and to be able to compare this value later on in the project to ensure that the image had not been changed  during analysis.

\emph{(6) Anonymise:} The created image is assigned a reference number. This is a randomised 6-digit number from  100,000 to 999,999.  The image files are stored on an encrypted USB hard drive.  A mapping is  retained until after the analysis phase to ensure an audit trail in case it is necessary to report a drive to law enforcement authorities.

\textbf{Analysis:}
 \emph{(1) Discover:} Each image is imported for file type analysis using a freely available tools such as Sleuth Kit's Autopsy.  A separate `case' is  created for each drive, using the forensic image ID as  identifier. 

\emph{(2) Tally:}
A tally of
 all file types is kept,  including  whether files are  encrypted or password-protected. 

\emph{(3) Carefully catalogue:} 
  Meaningful file names are noted to support content inferences.

\textbf{Post Analysis:}
\emph{(1) Validate:} Once  drives have been analysed,  the MD5 file hash of the forensic images is re-calculated and compared to initial values. This ensures that there drives have not been inadvertently altered.

\emph{(2) Eliminate Traces:} Mappings between drives and images will be destroyed.

\emph{(3) Securely Wipe Working Drives:}
 All drives are wiped to protect seller privacy and to prevent any subsequent deeper analysis of files. The drives are fully 7-pass formatted  to U.S DoD 5220.22-M (ECE) standard{\footnote{\url{https://www.media-clone.net/v/vspfiles/downloads/DoDEandECE.pdf}}}  using Apple’s Disk Utility Application. 

\emph{(4) Destroy Discarded Drives:} Any  drives found to be broken or unreadable are physically destroyed by removing the circuit board from the device and breaking the memory storage chip. 


\section{Analysis and Findings}\label{findings}
\paragraph{5.1 Survey \\}
We received 65 responses to our survey. Seven failed the attention check. After discarding these, we were left with 58 responses (M:49\%,F:44\%,Unspecified:7\%). 
64\% of our respondents had found a used USB drive lying around but 81\% said they themselves had never mislaid a USB drive. 


If they found a second-hand USB. 94\% would plug it i, 10\% would format it, 38\% would use forensics tools to check it out and 44\% would scan it with a virus checker. If the USB drive is a Rubber Ducky \cite{Ducky} or a USB Killer \cite{Killer}, their computers would be harmed. 

When we asked about encryption, 88\% claimed to know what encryption was, but only 17\%  actually used encryption. 59\% claimed to use password-protected USB drives, but this might well be a socially desirable response. 

\paragraph{5.2 Purchased Drives \\}
We purchased 122 drives, and had to discard 16. We were left with 106, of which 100 were analysed.
They ranged from 32MB to 256GB in size.
No viruses or malware were found on the drives.
Of the 100 analysed drives, only two had data that was immediately visible, i.e. sellers had made no attempt to delete the contents of the drive. When we examined all the drives using Autopsy, we found that 32 had no recoverable data, while the remaining 68 drives  had data that was recoverable to some degree, with full data recovery possible on 42 of the drives.

\begin{figure}
\centering
  \includegraphics[width=0.4\columnwidth]{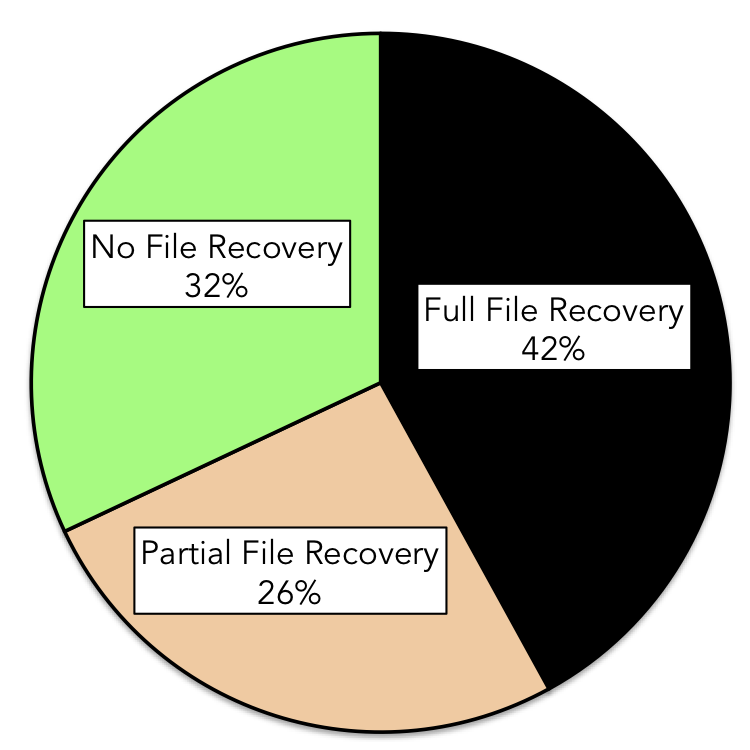}
  \caption{USB Drive Wiping Efficacy}\label{fig:drive}
\end{figure}

The cataloguing of the drives suggested a three-way categorisation of recovered files:\

\textbf{Low Sensitivity:} Evidence of downloaded videos; 
	Operating System Boot/Installation Drives (Microsoft Windows);
	Student Work;
	A range of promotional and press release materials.

\textbf{Medium  Sensitivity:} 	Image files with GPS location contained within EXIF Data;
	Personal photographs;
	Research studies containing clinical trials.

\textbf{High Sensitivity:}
   	 Image Files titled ``Passwords[1-10].jpg'';
	CVs, Personal statements, Employment contracts, Time sheets;
	Data relating to  apprenticeship trainees;
	Invoice records;
 Divorce information, Bank statements,  Health records,  and saved web pages.

\section{Discussion and Reflection}\label{discuss}

 The survey demonstrates that, at least for our respondents, 
public perception of  current recommendations is better than expected.  However, there is  evidence that  suggests that such awareness does not always convert to practice. 
 
There is potential for the  unintended consequences for those who left data on sold drives:
 (1)	In-person tracking down and harassment of the seller due to recovered photographs containing GPS co-ordinates stored within the EXIF data. 
(2)	Access to the seller's online accounts based on passwords obtained from image files (e.g. named `Passwords[1-10]').  If even one of these passwords were still valid,  there would be potential for financial compromise, especially if the seller had used the same password for more than one account.
(3) Identity theft via Tax Returns and Bank Statements.  If the account were  still active,  it might  be possible to siphon money from the account.
(4) Commercially sensitive information, in the form of invoices, makes it possible for a competitor to undercut the seller's prices, meaning that they could lose revenue.
Taking the above into account, and based on the fact that we found bank account records and tax returns, identity theft is a definite possibility.  If this happened, it could take them months, or even years, to resolve \cite{Ntuli}.


Whilst some of the findings cannot be directly compared due to our having categorised files in different ways, we have noted that Robins \etal \cite{robins2017investigation} detected that 230 of 271 drives (84.8\%), contained photographic images, whilst only  50\% of ours contained such files.
%
Robins \etal also found that 21 (7.72\%) of their drives had no recoverable data, whereas 32\% of our drives  had no recoverable data.
It is possible that awareness of the need to securely  wipe a USB drive prior to selling has improved. On the other hand, these samples are relatively small as compared to the number of drives sold yearly, and probably do not support definitive conclusions. 

Our findings suggest that current awareness drives are only  partly effective in raising the  general public's awareness of required digital hygiene. We could attempt to raise awareness even further with more effective campaigns. However, understanding the need to wipe drives will not change things unless people also know \emph{how} to wipe them properly. 98\% of the sellers in our study made some attempt to wipe their drives: the problem is that many did not know how to do it properly. 

There is also much that operating systems providers can do. Why do operating systems still persist in only wiping  file indexes in this day and age? Could the operating systems not detect that all files on a drive had been dragged to the recycle bin, and then ask the person if they want the drive  wiped securely? Let us now return to our two questions:



 \textbf{Q1 ---  Seller Risk:}    We discovered a great deal of information left on drives. 
 Using only  file names, we were still able to infer the kinds of data  left on drives. The sellers are probably unaware of the risks they run selling their used drives, and almost certainly unaware of the fact that the current `deletion' mechanisms  leave indelible traces that a clued-up person can recover. 
 This suggests that the seller risk is real, and significant. Moreover, it is also an indicator that the general public's digital hygiene is poor when it comes to this particular behaviour. 
 
 \textbf{Q2  --- Buyer Risk:}
We did not find any evidence that this was happening. 

\section{Conclusion}\label{conc}
The results from this study concur with outcomes from previous studies in 2009, 2011 and 2015. The drives we bought and analysed, seemingly owned by both individuals and organisations, were used to store highly sensitive and confidential information. Many  were insecurely wiped.
%

This situation almost certainly leads to privacy invasions that people are not aware of. Our survey suggests that whilst there is some level of awareness of encryption technologies, many do not convert awareness into practice. 
It is always worth doing more to raise public awareness. However, because reaching people is so hard, we suggest firstly that operating systems do a better job of wiping files, and secondly that these auction sites provide advice to sellers when they see a used device being listed for sale. By delivering just-in-time advice, we have the best possible chance of reaching the unwary, and preventing them from leaking their personal and sensitive data. 



\bibliographystyle{splncs04}
\bibliography{refs}
\appendix
\section{Appendix}
\subsection{News Stories}

\url{https://www.databreaches.net/thumb-drive-with-personal-information-of-milwaukee-employees-recovered/};
{\url{https://www.databreaches.net/nc-lost-thumb-drive-had-personal-info-of-5300-former-pitt-community-college-students/}};
{\url{https://www.databreaches.net/still-sending-data-via-unencrypted-thumb-drives-in-the-mail-it-will-cost-you/}};
{\url{https://www.databreaches.net/your-employee-info-is-in-the-mail-somewhere}};
{\url{https://www.databreaches.net/advancepierre-foods-sends-unencrypted-employee-401k-data-on-flash-drive-that-gets-lost-in-the-mail/}};
{\url{https://www.databreaches.net/ca-redwood-memorial-hospital-notifies-over-1000-patients-that-missing-thumb-drive-held-their-phi/}}\\ 
\ \\
Write Blocker:{{\url{https://www.cru-inc.com/products/wiebetech/usb_writeblocker/}}}\\
\ \\
Autopsy: {{\url{https://www.sleuthkit.org/autopsy/}}}

\subsection{eBay}
eBay’s User Agreement can be found at {\url{https://pages.ebay.co.uk/help/policies/user-agreement.html}}

\end{document}